\documentclass[twocolumn,preprintnumbers,amsmath,amssymb]{revtex4}
\usepackage{graphicx}
\usepackage{epstopdf}
\usepackage{amssymb}
\usepackage{dcolumn}
\usepackage{bm}

\begin{document}

\title{Bioengineering the biosphere?}

\author{Ricard Sol\'e$^{1,2,3}$ }
\affiliation{  
  \\ $^1$ ICREA-Complex  Systems   Lab,  Universitat Pompeu Fabra  -  PRBB,
  Dr. Aiguader 88, 08003  Barcelona, Spain
  \\ $^2$ Institut de Biologia Evolutiva, UPF-CSIC, Barcelona
  \\$^3$ Santa Fe Institute, 1399 Hyde Park Road, New Mexico 87501, USA
  }

\begin{abstract}
Our planet is experiencing an accelerated process of change associated to a variety of anthropogenic phenomena. 
The future of this transformation is uncertain, but there is general agreement about its negative unfolding that might threaten our own survival. Furthermore, the pace of the expected changes is likely to be abrupt: catastrophic shifts might be the most likely outcome of this ongoing, apparently slow process. Although different strategies for geo-engineering the planet have been advanced, none seem likely to safely revert the large-scale problems associated to carbon dioxide accumulation or ecosystem degradation. An alternative possibility considered here is inspired in the rapidly growing potential for engineering living systems. It would involve designing synthetic organisms capable of reproducing and expanding to large geographic scales with the goal of achieving a long-term or a transient restoration of ecosystem-level homeostasis. Such a regional or even planetary-scale engineering would have to deal with the complexity of our biosphere. It will require not only a proper design of organisms but also understanding their place within ecological networks and their evolvability. This is a likely future scenario that will require integration of ideas coming from currently weakly connected domains, including synthetic biology, ecological and genome engineering, evolutionary theory, climate science, biogeography and invasion ecology, among others.
\end{abstract}


\maketitle

\section{Introduction}

In a few human generations, our planet is likely to experience large-scale changes that will jeopardise the stability of our complex social and economic structures. Energy and demographic crises, biodiversity declines, increasingly frequent extreme events, along with water shortage and crop failure associated to climate change are already sending us warning signals (Scheffer et al 2001, Scheffer and Carpenter 2003, Scheffer 2009, Dawson et al 2011, Lenton 2011, Barnovsky et al 2012). We live in a time where the knowledge of our planet is greater than ever and the potential threads seem rather well defined. Scientists have depicted a grim perspective of our future. We are a major transforming force that is rapidly pushing our planet towards new, undesirable states. A consensus has emerged from climate science about a future, hotter planet that will make life difficult, if not simply incompatible, with a sustainable society (Lenton 2008). We have enjoyed a favourable window of 10.000 years, the so called Holocene period, where humans have been able to flourish as a dominant, creative and rapidly expanding species but also as a global geological force. The new human-driven era that emerges from the Industrial Revolution, the so called Anthropocene, is dominated by an increasingly obvious impact of human activities that are pushing the Earth outside its regulatory capacity (Steffen et al 2011).

	As it occurs with many other complex systems (May 1977) continuous changes in parameters that control the state of given system often end up in catastrophic shifts once tipping points are reached (Scheffer 2009, Sol\'e 2011, Hughes et al 2012). This is the case of the average concentration of carbon dioxide: once some critical levels are reached, our current climate state is likely to be replaced by another global pattern resulting from a runaway greenhouse effect (Solomon et al 2007, New et al 2011). A macroecological analysis of energy use and economic activity also indicates that the current tendency might end in a social and economic collapse (Rockstrom et al 2009). Similarly, many ecological systems will face rapid declines towards degraded and even bare systems with no species left (Suding et al 2004). This is illustrated by arid and semiarid ecosystems (Rietkerk and van de Koppel 1997, Scanlon et al 2007, K\'efi et al 2007, Sol\'e 2007) where warming, steady declines in rainfall and increased grazing will trigger rapid changes towards a desert state and are specially vulnerable (Thornton et al 2011). Evidence for such sudden changes exist, as shown by the shift from a green Sahara to the current desert state, which took place 5500 years ago (Foley et al 2003). Rainforest ecosystems, reefs and boreal forests might also face serious declines (Barnovsky et al 2012, Hughes et al 2013). In some cases, as illustrated by the collapse of fisheries, they have already occurred while the awareness and reactivity of society to such sudden loss has been far from optimal (Scheffer et al 2013).

Many studies have addressed possible ways for remediating these potentially catastrophic situations. Humans too have been effectively operating as ecosystem engineers (Vitousek et al 1997) by adapting the biosphere to their needs, while expanding their populations in a hyper exponential fashion. Because our long-term influence, vast amounts of energy-intensive fossil fuels have been used to power our civilisation, reinforced by the accelerated growth of agriculture from the Neolithic revolution. Profound alterations of the water and nitrogen cycles are a direct consequence of these unsustainable practices. Moreover, an ongoing rearrangement of biotic systems has been taking place, mainly due to habitat loss and biological invasions (Elton 1958, Drake et al 1989). By doing that, we are changing the face of our biosphere, placing ourselves close to a planetary-level critical transition. Can the situation be reverted?

Existing approaches, to be summarised below, include reforestation, geo-engineering and emission cuts, among others.  However, 
the scale of the problem, the staggering economic costs and its accelerating pace constitute a major barrier to restore previous states in a sustainable way (Folke et al 2011). Moreover, we need to face the nature of our biosphere as a complex adaptive system with multiple interacting species, nonlinear responses, complex feedbacks and self-organizing patterns (Levin 2002, Sol\'e and Levin 2002). There is a strong 
asymmetry between cumulative anthropogenic impacts and our slow and limited capacity for counterbalancing them on time. Such asymmetry implies that we might have a narrow time window to properly react to the challenge. In this paper I suggest a rather different approach, which requires an engineering perspective, grounded in the design of modified life forms and intervention. But, above all, requires a new merging of disciplines, particularly at the unexplored boundaries between synthetic biology and ecological theory. Because it requires humans as agents for EarthÕs transformation, the remediation strategies suggested here imply a modification of natural ecosystems. This is, no doubt, a controversial matter (Callaway 2013). The advantages and drawbacks of this approach, along with implementation strategies, are outlined below.

\section{Terraforming Earth?}

	Restoring a sustainable EarthÕs state necessarily requires to confront the scales of space, time and energy on the planetary level. That means that whatever the solutions found, they go beyond any human standard engineering scale. Before looking at our own biosphere, let us first make a turn by considering the other single scenario where such engineering problem has been proposed, namely the problem of "Terraforming" Mars (McKay et al 1991). The idea is, in a nutshell, to introduce artificial modifications that trigger a runaway process capable of displacing the planet's state towards a new steady state with higher temperatures, water levels and thicker atmosphere. That could be achieved through the use of greenhouse gases (Lovelock 1988) although at very high costs. It would be also achievable or by means of appropriate microorganisms (Rothschild and Mancinelli 2001) capable of adapting and growing under extreme conditions. In both cases, a relatively small perturbation is expected to get amplified, ultimately affecting the planetÕs geochemical cycles. The first possibility is unlikely to be feasible due to the associated costs. But the use of extremophiles, such as some bacterial species of Carnobacterium (Rothschild and Mancinelli 2001, Nicholson et al 2012) have been shown to tolerate extreme conditions (including low pressures and temperatures along with anoxia).

	In this paper we will use the previous scenario as a starting point to discuss how the release of genetically manipulated organisms could be used to restore habitat and climate unbalances at local, regional and even global scales. Such possibility has not been raised before. Instead, within the context of global warming, existing proposals consider geoengineering (Lovelock and Rapley 2007, Schneider 2008, Vaughan and Lenton 2011, Caldeira et al 2013). In contrast with reduction of emissions, this climate engineering scheme (directed to mitigate global warming) operates directly on diverse physical or chemical factors. The cost of most proposed solutions is typically enormous, as a consequence of the massive scales involved. These solutions include a broad variety of possibilities, from hundreds of thousands of towers to capture carbon dioxide to trillions of small, free-flying spacecrafts (Vaughan and Lenton 2011, Caldeira et al 2013). Lower costs but high risks are expected from using aerosols, to be injected in the stratosphere to counterbalance greenhouse gases (Lovelock 2008). Other strategies, such as iron seeding to trigger plankton blooms have failed to meet their expectations. Even despite the limitations of these proposals, a common message is that the price of not preparing for the future will be much higher than the investment in any of the previous possibilities (Schneider and Mesirow 1976).

	How to deal with the large scale problem that we face here? If geoengineering is not the right approach, what can be the alternative? We should look for feasible solutions capable of (a) solving the scale problem at a reasonable cost, (b) restoring the desired systemÕs state over an appropriate time scale and (c) minimize the risks of undesired evolutionary dynamics. The approach suggested here is that such solutions might soon exist at the crossroads between ecosystem engineering (Odum and Odum 2003) and different approaches oriented towards engineering living systems, particularly synthetic biology (Drubin et al 2007) and genetic engineering of plants (Mittler and Blumwald 2010). So far, all these approaches have been developed within a lab or farm context where containment is a major concern (Church 2005, Dana et al 2012). Not surprisingly, biosafety issues related to the potential release of engineered organisms or genetic material have become part of the research agenda. Given all the unknowns, containment has been at the centre of these disciplines as much as their design principles. What I want to suggest is an orthogonal, but may be complementary: ÒTerraforming EarthÓ by engineering new synthetic organisms capable of counterbalancing undesirable trends. A major difference of this type of engineering is obvious and crucially departs from geoengineering: since living entities self-replicate, an engineered organism capable of large-scale dispersal would eventually reach, by growth and reproduction, the desired scale. This could be achieved within reasonably short time scales and the proposal is not limited to capturing carbon dioxide: as an example, engineered bacteria could be designed to help plants facing stressful habitat conditions in order to improve their survival, perhaps enhancing desirable soil microbial communities. Other manipulations affecting photosynthetic efficiency or light-sensing properties could also change the ways we can repair damaged habitats (see below). 

	The release of a living system that has to spread over large biogeographic areas should be considered cautiously (Snow et al 2005, Pilson and Prendeville 2004). How they can affect community-level traits requires a multi-scale view of ecosystem processes (Whitham et al 2006). We already know that a harmful invasion of a given community from an engineered species (Sanvido et al 2007) as it occurs with non-engineered ones, is difficult, since multiple barriers need to be overcome (Blackburn et al 2011). We also need to consider that, given the fast progress and cost reductions associated to this technology,  it is not too soon to start exploring the set of problems presented here. By its nature, it requires the merging of multiple disciplines and a serious consideration of the tradeoffs between designed forms of life and the evolutionary responses to their introduction. However, a well developed theoretical framework already exists concerning the reliability of ecosystems and the role played by key factors such as species redundancy (Pimm 1991, Naeem 1996). Moreover, it can be argued that there is only one Earth-like planet that we can study and we can easily conclude that the lack of alternative scenarios makes the whole proposal highly speculative. However, as discussed below, not only one, but many case studies might actually be available to us, and much closer that one would expect. 

Several differences can be noticed while comparing the Mars Terraforming scenario and the one considered here. Above all, 
Earth is the planet we live on. Mars requires a bottom-up development of a resilient network of biotic-atmospheric interactions enhancing life. That means a sequential process of niche construction, where cells adapted to the extreme conditions of the new planet must be capable of modifying this environment in order to grow in increasingly more efficient conditions. Under these conditions, evolutionary dynamics is on our side: selection for more efficient metabolic pathways, better protection mechanisms against radiation and climate extremes would spontaneously trigger improvements. In our planet, bioengineering would be a more top-down strategy, since the network of existing species and their biogeochemical context is already established. Engineering new species means to redefine the existing network of interactions so that we can restore previous steady states or perhaps create novel ones. That would require building new symbiotic relations with existing species and considering several facets of the synthetic one, from efficiency to evolvability.

\begin{figure*}[htbp]
\begin{center}
\includegraphics[width=14 cm]{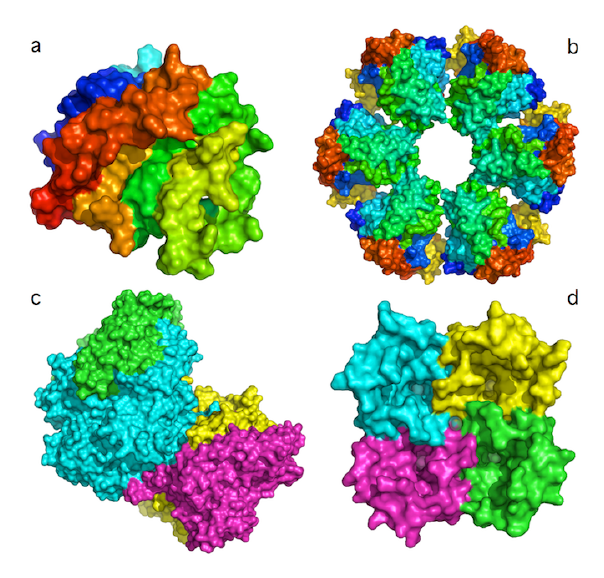}
\end{center}

\caption{
Engineering of molecular structures allows to overcome limitations to metabolic efficiency, water shortages and nitrogen fixation. Here we show the molecular structures of (a) phytochrome, used by plants as a light detector molecule, which has been modified in such a way that photosynthetic processes and growth can take place under shading. In (b) we display the Rubisco complex of Tobacco, the key enzyme associated to CO2 assimilation. By knocking out the gene associated to this enzyme and introducing the one from a cyanobacteria species, higher efficiency was achieved. A third example is the enzyme nitrogenase (c) which is used by Bacteria and Archea for nitrogen fixation. In (d) we display the Aquaporine 1 transporter protein, which can help improving water use efficiency. The 3D pictures have been created using the Pymol software package for molecular visualization (http://www.pymol.org).
}

\end{figure*}

\section{Synthetic ecosystems}

 The proposal described here originates from the assumption that a synthetic organism can act, in some circumstances, as ecosystem engineer (Jones et al 1994) capable of modifying the existing energy balances and/or nutrient flows. In nature, bacteria and microscopic algae in particular have played a major role in shaping our EarthÕs climate (Kasting and Siefert 2002, Falkowski et al 2008, Lenton and Watson 2011) and could help us restore lost balances. Such interaction might have created a stable homeostatic robustness, as formulated in LovelockÕs Gaia hypothesis (Lovelock 1992, Lovelock and Margulis 1974, Lenton and van Oijen 2002). This view suggests that negative feedbacks between biotic and non-biotic compartments are affected by close-loop controls in such a way that the biosphere would maintain itself 
 within the range of parameters facilitating life. Other views suggest that positive, destabilising feedbacks might be 
 no less important (Ward 2009, Field et al 2007). If designed organisms have to be released, they should act preventing positive feedbacks and instead creating or reinforcing existing close-loop controls. 
 
Several potential situations and strategies can be considered, depending on the scale and nature of the problem. 
A given environmental variable might need to be restored by means of a modified microbe or engineered plant species, which will act (effectively) as a designed invader. On a small or regional scale, this perturbed system can be a contaminated terrain or a degraded semiarid land. On the largest scale, reducing greenhouse gases or nitrogen excess would define two major problems. Concerning the first scenario, bioremediation strategies have been already addressed since the mid 1980s mostly associated to organic pollutants or heavy metals, with different degrees of success (Cases and de Lorenzo 2005, de Lorenzo 2008). A major obstacle to evaluate their potential in the field is, of course, related to biosafety issues. New methods from synthetic biology are changing the landscape  (Sayler and Ripp 2000). The area has been useful to understand the interplay between metabolic-level constraints and ecological-level (population) factors. Many problems arose when using strains of modified bacteria due to poor responses to environmental stress, reduced selective advantage associated to accumulated waste products and unpredictable factors resulting from a poor understanding of physiological traits. In other words, modified genetic and metabolic networks where not enough: as it occurred with PCB-degrading strains with all the genes required for it but nevertheless failing to do the biodegradation as expected. Predictive power has increased thanks to genomic search, along with systems and synthetic biology approximations (Schmidt and de Lorenzo 2012). 

	Synthetic biology represents the last step in our potential for modifying natural systems. The field has been growing fast in the last decade and has emerged as an engineering approach to modify or even create de novo living cells tissues and organs (Purnick and Weiss 2009, Khalil and Collins 2010) and is considered a promising new framework capable of facing a large array of fundamental problems, including new therapies, development of drugs or biofuel production (Wang et al 2013). Different approaches have been taken to modify living entities or even create new ones from scratch (Sol\'e et al 2007a, 2007b). Among them, two main paths are being followed. One is the top-down approach, where we start from an already existing species and modify it. This path is being followed by research projects involving the reduction of genomes and the concept of a minimal gene complement (DeWall and Cheng 2011). Minimal genomes are characteristic of both free-living and endosymbiotic species (McCutcheon and Moran 2012) but free-living microbes can also exhibit a reduced genome, as illustrated by the cyanobacterium Prochlorococcus, one of the most abundant marine microbial genus (Dufresne et al 2003). Despite not as well developed as with Escherichia coli, libraries of genetic constructs (so called biobrick parts, see Wang et al 2012) for cyanobacteria are already available (Baker et al 2006). Although most efforts in this area have been directed towards the synthesis of biofuels and other chemicals, improved CO2 fixation and light harvesting have also been achieved (Ducat et al 2011).  In this context, a widespread microorganism containing a minimal genome appears to be an optimal candidate towards engineered carbon sequestration in the ocean.  

	The synthesis of a whole bacterial genome dramatically showed that a large-scale, genome-level engineering is feasible (Gibson et al 2008) although we need to accept that little is really known about how genes actually interact in a whole network. Genome reduction is still under development but it will surely deliver reliable designs in a near future (Esvelt and Wang 2013). The bottom-up approach to minimal cells includes the creation of protocellular systems (Szostak et al 2001, Rasmussen et al 2008, Luisi 2006, Sol\'e 2009) up from pure chemistry, not necessarily the biological one, thus requiring to cross the twilight zone separating living from non-living matter. This achievement has not been successful so far, but it is not unlikely to happen in a near future. A great advantage of this approach is that designed protocells might be more easily controlled. They can even be much less constrained in their evolvability, while be very efficient in performing a given metabolic function over a range of conditions (Zhang et al 2008). This possibility has been discussed within the context of large-scale bioremediation strategies An example is provided by the suggestion of using large artificial limestone reefs acting as the physical substrate for synthetic protocells that would be programmed to improve water quality in contaminated or oxygen-poor aquatic ecosystems. Although still a speculative arena, tentative ways of using protocellular constructs combined with artificial reefs have been outlined (Armstrong and Spiller 2010). In these still speculative case studies, both the reef and the microbial population would grow and self-regulate each other. 

	The use of plants (mostly crops) as the targets of engineering complements the previous single-cell species scenarios. Crop development, selection and geography expansion has been by far the largest ecological engineering process performed by humans. Plant domestication led to highly enhanced yields and a revolution in human history, powering the exit from the Holocene (Diamond 2002). Nowadays, genetic engineering techniques have enormously accelerated the potential for rapidly modulate endogenous metabolic pathways through multi-gene transformation (Palumbi 2001, Zorrilla-L\'opez et al 2013).  Similarly,  sensible improvements to metabolic efficiency and environmental sensing have been obtained (figure 1a-d). Alternative carbon-fixation routes (distinct from those associated to the Calvin cycle) have been elucidated and offer new ways to improve them (Farre et al 2014).  Recent work has shown that it is possible to transfer the genes from cyanobacteria coding for the Rubisco enzyme (probably the most abundant protein on Earth) into plant crops (Ducat and Silver 2011) leading to higher rates of carbon fixation (Lin et al 2014).

Another breakthrough of molecular engineering is the modification of plant phytochrome (Whitney et al 2011). This light-sensing molecule is a crucial component of plant physiology, acting as a light detector that triggers plant responses to given levels of sunlight. Decreased radiation input leads to a less active state. However, that threshold response could be tuned, in principle, by changing a molecular switch (Burgie et al 2014). A third example involves the enzyme nitrogenase, which has been used to improve nitrogen fixation in cereals. Although molecular nitrogen accounts for the largest fraction of atmospheric gases, it cannot be directly fixed by plants. Instead, this limitation is circumvented in agriculture by chemical fertilisers, resulting in greenhouse gas production and damage to aquatic ecosystems through eutrophication as well as other environmental problems (Canfield et al 2010). The possibility of designing synthetic plant-microbial consortia capable of fixing nitrogen would represent a major advance. Similarly, advances in the Crassulacean acid metabolism (CAM) suggests that water use efficiency in arid ecosystems could be improved (Borland et al 2011). This can be a specially crucial approach to reduce the future impacts of water depletion (Carnicer et al 2011). Current strategies include engineering key molecules as the CO2-transporting aquaporin (Sade 2014) or moving CAM into C3 crops. A different proposal is based in engineering of bacteria capable to accelerate plant root development by means of hormonal stimulation (\url{http://2011.igem.org/Team:Imperial_College_London}). In this case, the microbe would secrete auxin (a plant hormone) which would have been previously coated seeds before planted. By enhancing a more efficient establishment of roots, land erosion could be prevented while soil moisture could be increased. 

	This paper does not systematically consider all the alternatives. I have not mentioned viruses, for example, as potential targets of bioengineering designs. One reason for this exclusion is that viruses are the most rapidly evolving part of our biosphere and thus genetic firewalls might face serious challenges. However, they are also a major driver of global geochemical cycles (Wilhelm and Suttle 1999, Fuhrman 1999) strongly influencing ecological processes in the ocean (Suttle 2005). Moreover, our view of viruses as parasites or pathogens has been shifting over the years as they play a beneficial role in many different associations with species belonging to all kingdoms (Roosinck 2011).  Since mutualistic links might constrain their evolvability, engineered viruses should not be discarded. Similarly, other symbiotic relationships could also play a role once we fully develop the right engineering tools. This is the case of synthetic chloroplasts (Agapakis et al 2011) by introducing engineered photosynthetic bacteria within animals. Similarly, we should not exclude the use of xenonucleic acids (XNAs) where novel informational biopolymers are used (Schmidt 2010). 
	
	Unconventional ways of strongly limiting spread and evolution can also be obtained by using cell consortia (Brenner et al 2008) where different cells involving different engineered parts perform together a given function, including cooperation (Shou et al 2007). A non-standard approach has been shown to provide a source of modular design while strongly departing from both biological and engineering principles (Regot et al 2011, Macia et al 2012). In this so called Òdistributed multicellular computationÓ, the requirement that all parts must work altogether automatically places constraints to evolution and spread. Moreover, this class of so called cellular computing (Amos 2004) is grounded in a design scheme that does not follow standard engineering approaches. It is instead closer with the ways biological systems manage information in a distributed fashion (Sol\'e and Macia 2013).

\begin{figure*}
\begin{center}
\includegraphics[width=16 cm]{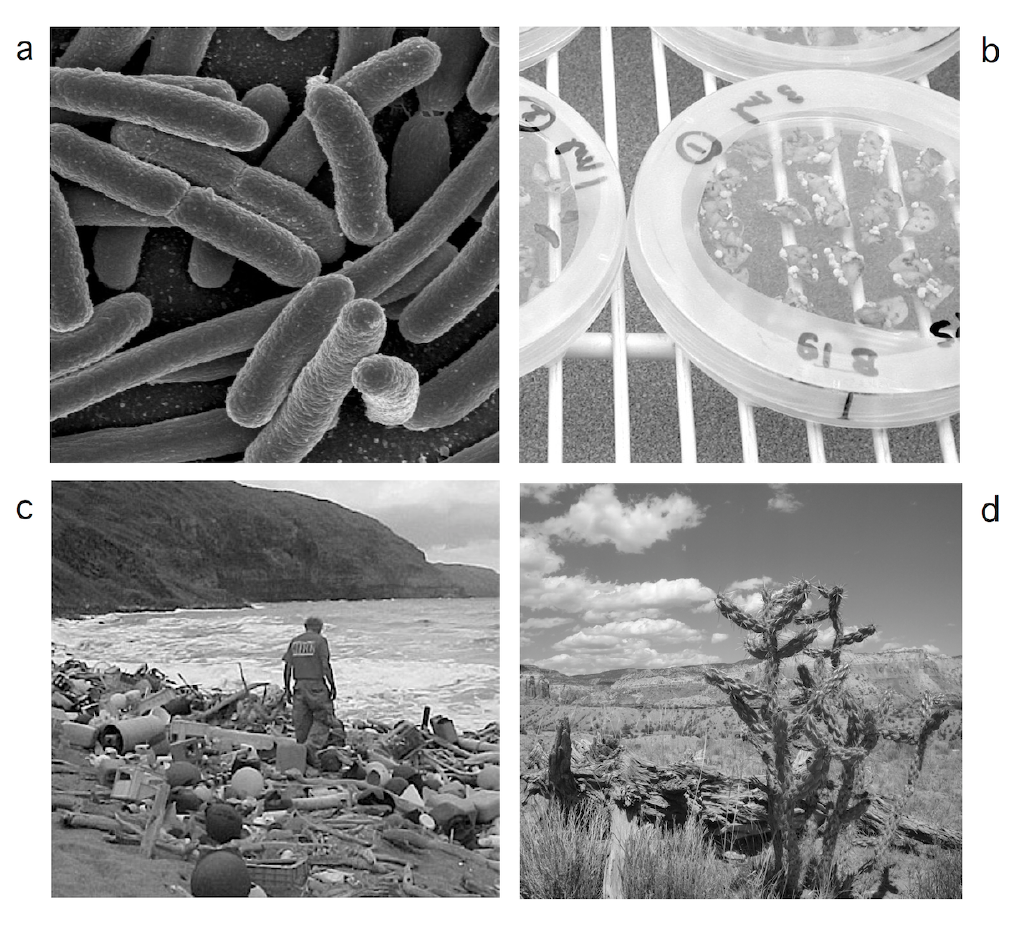}
\end{center}
\caption{
Bioengineering existing ecosystems will require a proper choice of the kind of model organism to be engineered. 
Potential organisms could be for example (a) bacteria (here we show E. coli) that could be 
programmed to perform specific functions or (b) engineered plants (here genetically transformed Solanum plants). 
Moreover, very different scenarios can be considered in relation to where these modified organisms can be released. This can include man-made 
substrates, such as (c) marine debris and also (d) arid an semi-arid ecosystems (picture 
courtesy of S. Valverde). In both cases, a combination of genetic firewalls and habitat constrains would help to contain engineered organisms. }
\end{figure*}

\section{Predicting the outcome}

In order to address the problem of accidental release of modified organisms, protocols for contention appeared as soon as genetic engineering started to develop. However, the early claims of the impact of recombinant DNA technology have been shown to be largely unfounded (Berg and Singer 1995). Synthetic biology has raised similar concerns  and different biosafety protocols have been established. In fact, the design of new organisms that perform given functions can include, as part of the new circuits, control components facilitating biocontainment (Wright et al 2012). Metabolic dependencies, programmed cell death, auxotrophic constraints or strict synthetic symbiosis are well known ways of implementing genetic safeguards. In this area, important advances in genetic engineering have been achieved. These include introducing switches that activate once a given external signal moves below a threshold or conditional suicide (Moe-Behrens et al 2013). Most of these designed mechanisms have been build using prokaryotic species, were higher mutation rates favour evolutionary responses and both failure of the switch and spread of undesirable genetic material are possible outcomes. However, recent work suggests that it might be possible to engineer the evolutionary potential of synthetic organisms (Renda et al 2014). Further improvements and genetic firewalls will be achieved by using eukaryotic species, particularly by working on microbial species with minimal, well characterised genomes as well as plant crops and engineered microbes associated to them (figure 2a-b). But we also need to face a reality: hacking living systems is becoming a cheap and widespread task, as standardisation of genetic parts becomes a reality (Endy 2005, Ledford 2010, Schmidt 2008). Little is known about the potential for survival of these engineered strains nor what ecological-oriented biosafety measures should be used. 

	Released synthetic organisms can remain locally established when the chosen environment and the habitat constraints make spread to other areas unlikely (figure 2c-d). This can be the case of semiarid ecosystems, which may be largely responsible for CO2 emissions (Poulter et al 2014). Here the strong limitations to growth due to water shortages and poor soils are in fact a challenge for any introduced species. Because of these strong niche limitations, containment might be simpler than expected, while the overwhelming amount of solar power should be an advantage. Since the most likely scenario might involve manipulating microbial-plant interactions, the mutualistic loop would also help defining a controllable scheme. Beyond natural systems, we can also consider existing structures resulting from anthropogenic activities as a likely substrate for bioengineering. I would suggest two of them: cities and plastic marine debris. Cities are the greatest hot spots of carbon dioxide emissions, hosting more than half of the global population (Grimm 2008).  Urban areas represent an enormous substrate for growing and controlling engineered organisms. Urban landscapes represent a major advantage, namely a constant supervision of the synthetic species behaviour, helping in a constant supervision of their growth and should coexist with current efforts of management of urban ecosystem services (Andersson et al 2014). In this context, ongoing work has been done in exploring different ways of using modified organisms in a urban-related context. An example is the BacillaFilla project (\url{http://2010.igem.org/Team:Newcastle/problem}). Here the idea is using modified strains of Bacillus subtilis to repair concrete deterioration. These cells would move inside the cracks, filling them with calcium carbonate and other components such as levan glue (a polysaccharide). Once the filling process ends, bacteria die and become part of the biologically grown material. In this way, we would have a Òperform-and-dieÓ scheme of intervention. Since concrete is a major component in construction and its industrial production is associated to a number of environmental problems, a bioengineering strategy that can considerably extend the material lifetime could have a considerable impact. 

	Secondly, the so called ÒPlastisphereÓ (Gregory 2009, Zettler et al 2013) provides another anthropogenic (but human-free) substrate that can be exploited as a niche for synthetic organisms. Plastic debris has been extensively colonised by marine life forms from microbes to bryozoans, hydroids or molluscs (Barnes 2002). An example of this could be the use of plastic oceanic vortices, where large amounts of long-lived polymers have been accumulating over the 20th century at an accelerated pace, although recent work revealed a much smaller amount of debris than expected (C\'ozar et al 2013). This actually suggests that there might be already some biotic degradation process at work. Waste reservoirs, in general, constitute a threat but can also be an opportunity to provide the appropriate habitat for synthetic microorganisms. An imaginative bioremediation proposal (\url{http://2013.igem.org/Team:Imperial_College/mainresults}) is to engineer adhesion between existing marine bacteria-forming biofilms so that small plastic particles attach to others, forming larger clumps or even plastic islands. The existing polymer habitat could also be used to create a diverse consortium of engineered, cooperating bacteria associated to polymer degradation coupled to enhanced carbon fixation. As a relatively confined spatial system, it would help to monitor the spread and efficiency of the bioengineering process and how robust its development over time. 

	How can we predict the possible outcome of manipulated life forms in a complex biosphere? What might be the most reliable and safe design to be developed? One avenue should involve the classical approach taken in ecology of using micro- and mesocosm experiments, which have been extensively used in ecological engineering (Odum and Odum 2003, Stewart et al 2013). These are spatially confined, carefully controlled laboratory habitats where temperature, humidity and nutrient intake are tuned, while the ecosystem responses are monitored. Many experimental protocols associated to plant engineering match this description. Additionally, mathematical and computational models should be used, including available information and capable of making forecasts across multiple scales (Woodward et al 2010, Evans 2011). How evolution of the engineered 
	systems occurs in these small-scale environments is also a very important topic that should be addressed (Peisajovich 2012). Another place to address our previous questions lies inside us, in the so called human microbiome (Huttenhower et al 2012). Our microbiome accounts for the vast majority of genetic information carried by our bodies: more than three million genes are carried by the microbiome species, to be compared with the estimated twenty thousand human genes (Ackerman 2012). Trillions of microbes inhabit our gut, skin and other organs and are clearly associated with health and disease (Cho and Blaser 2012) acting as a regulator of their host physiology. The microbiome 
is linked to a wide variety of diseases, including those related to metabolism, autoimmune responses and cancer. Not surprisingly, synthetic biology has become a major potential path to harnessing these naturally commensal microorganisms to prevent infections, deliver desired molecules targeting given diseases (Ruder et al 2011). 
	
The host-microbiome network of interactions is the outcome of a long co-evolutionary process and seems to be essential in maintaining organismal homeostasis (Dethlefsen et al 2007). It defines a complex web of interacting species, whose links are strongly influenced by ecology and geography (Smillie et al 2011). Perhaps not surprisingly, the biodiversity patterns displayed by microbiomes have many things in common with the regularities found in species-rich ecosystems. More importantly, a proper understanding of both the healthy and diseased microbiome requires an ecology-level perspective. In this context, it has been proposed that processes associated to invasion by pathogens can be understood in terms of standard invasion ecology. Treatments and recoveries from disturbance can actually be represented in terms of shifts among alternative states (Costello 2012, Pepper and Rosenfeld 2012). The human microbiome can thus be regarded as a large scale, multispecies ecosystem where both the effects of perturbations and potential recovery scenarios can be traced in detail.  

	Wether synthetic microorganisms are delivered as free entities, to become part of soil microbiomes or as integrated, symbiotic species within more complex hosts, it will be important to predict the success of these new species as they become part of their new environments and start expanding. This is a major topic within invasion ecology (Elton 1958, Parker et al 1999, Simberloff and Rejm‡nek 2011, Strayer 2012) where a key question is what makes a new species getting established within a community (Vitousek et al 1996). Predicting invasion success has been addressed in many different ways, using different sources of information, from network structure to morphological traits (Romanuk et al 2009, Azzurro et al 2014). Moreover, invaders can succeed only temporally, sometimes collapsing into extinction (Simberloff and Gibbons 2004) despite an initial successful population expansion. 
	
	Since we might be interested in a transient response where the engineered species performs a function, such so-called boom-and-bust cycles might be the appropriate dynamical targets. It is also worth noting that the initial impact of invaders can decline in the long term, with species richness and productivity restored after a few decades (see for example Dost‡l et al 2013, Strayer et al 2013). A rule of thumb is that a successful invader has to be capable of occupying some available niche space not exploited by the members of the receptor community (Shea and Chesson 2002). Are there equivalent rules of thumb for bioengineered systems? Future developments towards a theory of synthetic invaders will require to address the problem of how to define the niche requirements of the designed species in order to succeed while preserving their host habitats. In this context, it should be also noticed that some invaders have been shown to have positive effects as ecosystem engineers, facilitating native fauna instead of harming it (Wright et al 2014).

\begin{figure}[htbp]
\begin{center}
\includegraphics[width=8 cm]{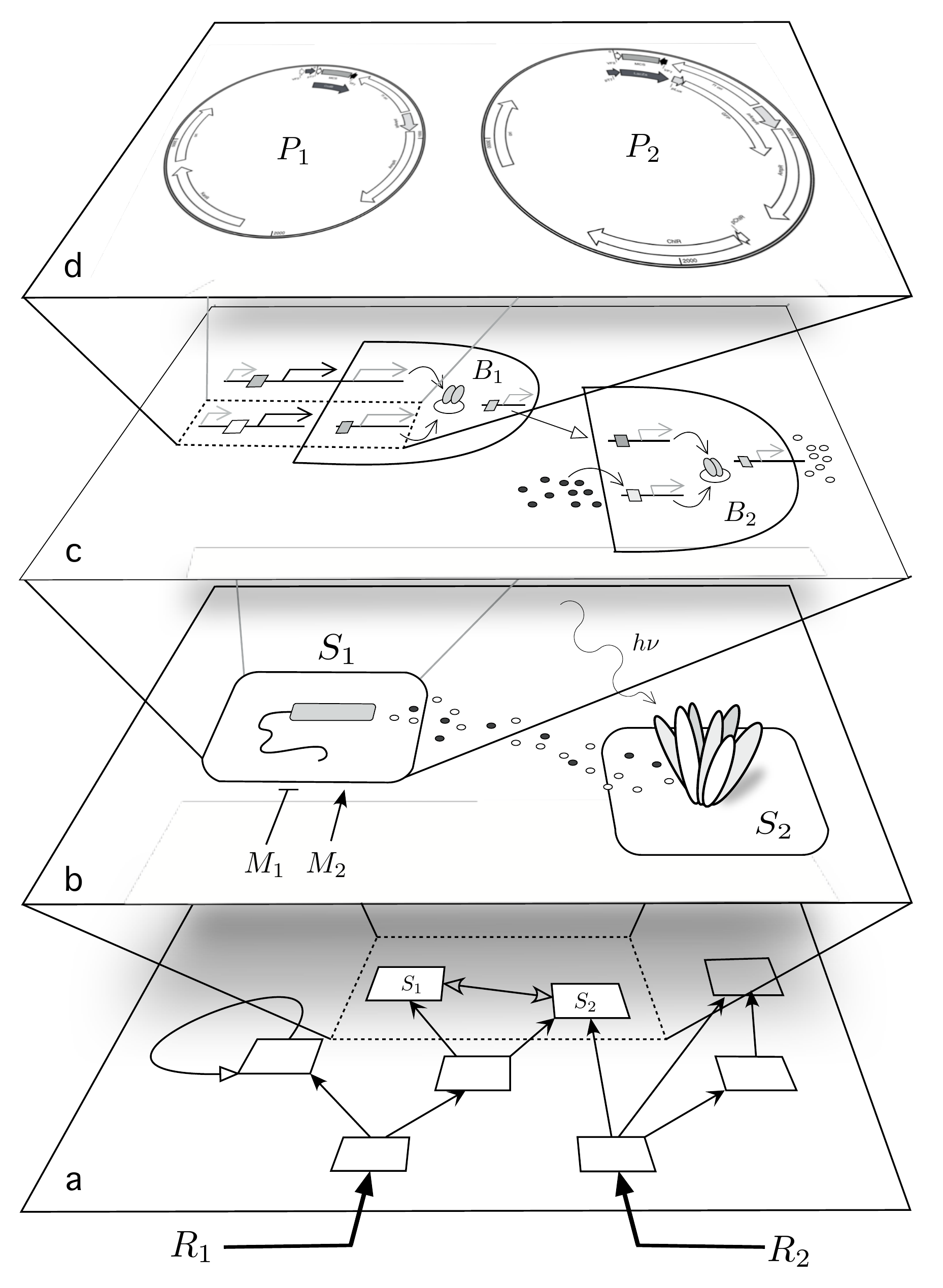}
\end{center}
\caption{
A systems view of bioengineering ecosystems using synthetic biology. Multiple scales need to be considered in order to design a given functionality to be 
embodied within a given ecological web. From bottom to top, we have (a) the ecosystem level, where the web of interactions among species ($S_k$) and 
external inputs and resources ($R_k$) needs to be considered, along with biogeochemical constraints, (b) the species interaction level, which can involve 
multiple kingdoms and different kinds of relationships, particularly mutualistic ones, and where a given species might be affected (modulated) by external 
signals $M_k$), (c) the single-species design circuitry to be modified by genetic engineering of specific logic blocks ($B_k$) and (d) the molecular toolkit (which 
involves, for example, molecular constructs on plasmids, here indicated as Pk) used to engineer the specific molecular machinery and regulatory 
interactions.
}
\end{figure}
  
 \section{Conclusions and discussion}  
   
   What can be the impact of terraforming our own planet? Should we even consider that possibility? Can we deal with the complexity associated to such scenario? Some have compared geoengineering approaches to climate remediation with the Manhattan project (Michaelson 1998) but I think the right dimensions in terms of the challenge are better met with the proposal outlined here.  An intervention that can modify the biosphere in controlled ways to reach a new steady state compatible with a planet where humans can live should be seriously considered and joint international efforts might be required at some point. Beyond the technical challenges (that might require a war-level effort) and the much needed theoretical basis, the problem of how decisions should be adopted at the regional and global scales, and whether consensus is even achievable, is part of the scenario presented here. As it occurs with geoengineering (Schneider 2008) global cooperation might be required. Nevertheless, what seems clear is that the tools for building some of the candidate organisms discussed above will be available sooner than later and not necessarily coming from academic institutions (Church and Regis 2012, Porcar and Peret\'o, 2014). A Òdomesticated biotechnologyÓ, as defined by Freeman Dyson (2007) implies an enormous combinatorial power. We should not wait to start thinking about the ecological effects of new organisms as potential invaders. Instead, designs and strategies favouring successful, but limited, establishment should be started to be tested as soon as possible.

	We cannot foresee all future changes that will unfold in the next decades as a consequence of our impact. What we do know is that all sorts of evidence point towards an unsustainable outcome where our society is likely to fail unless serious measures are taken (Brown et al 2011). Novel approaches are needed, and often the novelty emerges at the crossroads between apparently distant fields. Nowadays, synthetic biology and ecological theory are loosely connected, but it is at the intersection between these two major disciplines where some solutions might reside. The essential message is that we need to reinvent a small part of nature in order to preserve as much as possible while we guarantee our persistence as a species in a sustainable way. To reconnect with the biosphere, we might need to redesign it using a multiscale, complex systems view of ecological systems (Figure 3). In this context, useful lessons can be extracted from controlled field experiments involving species additions or removals and the consequent trophic cascades (Brown and Heske 1990, Estes et al 2011).  Similarly, a variety of model approaches will be needed, from simple population-nutrient flow models involving differential equations (DeAngelis, 2013) to large-scale models of climate and bio-geochemical cycles (Lenton et al 2007). The later would be useful to explore the impact of tentative strategies of carbon sequestration, providing some clues concerning the expected C:N:P stochiometric ratios resulting from bioengineering. Here too some basic models (Klausmeier et al 2004) and a proper choice of target model organisms among the available functional diversity (Arrigo 2005) will be essential to develop this framework.

	The fact that we are proposing an engineering perspective does not mean that we can change ecosystems in whatever way we wish. As ecologist E. O. Wilson stated, this is Òone planet, one experimentÓ (Wilson 2010) and our first obligation is to scientifically evaluate any artificial modification scheme. No one should get the message that we Ògive up on the BiosphereÓ.  Instead, we need to improve our knowledge on how complex ecosystems work, while considering possible interventions and adopt them under international agreements. A better understanding of how and ecosystems decline is needed in order to drive future research in this (still to be defined) field. In this context, the study of novel ecosystems resulting from human action, climate change and other accidental or deliberate events is by no means new (Hobbs et al 2006, Seastedt et al 2008, Hobbs et al 2009, Lurgi et al 2012). Lessons from past species invasions and their long-term effects should actually guide us in extracting useful lessons and some cautionary tales (Strayer et al 2006, Sax et al 2007). 
	
	We can use this knowledge to intervene in ways that preserve both biodiversity and human well-being. Some key ideas from a complex systems view of ecosystems should guide us (Levin et al 1997, Levin 2000, Sol\'e and Bascompte 2006) and our design efforts must be driven towards bioengineering reliability (Holling 1973, Peterson et al 1998). Technological solutions must be developed in parallel with strategic decisions including sustainable growth, a proper use of energy and material resources and species conservation. Unless we maintain all parallel efforts to slow down our impact on the Biosphere, no safe way out from the Anthropocene will exist. The challenge ahead is enormous and the scenario presented here must not be taken as a free lunch view of ecosystem remediation based on a blind faith in the success of technology. Instead, it should be seen as a rational framework to help escaping from an ongoing runaway effect, at least temporally. If anything, we certainly are running out of time.

\vspace{0.75 cm}
{\bf \large Acknowledgments}
\vspace{0.25 cm}

The  author would like  to thank the members of the Complex Systems Lab and many colleagues at the Santa Fe Institute for fruitful discussions. Special thanks to Daniel Amor, Ernesto Azzurro, Steve Carpenter, Victor de Lorenzo, Salvador Duran-Nebreda, Santiago Elena, Harold Fellermann, Joaquim Garrabou, Tim Lenton, Simon Levin, John McCaskill, Raul Monta–ez, Melanie Moses, Manuel Porcar, Juli Peret\'o, Stuart Pimm, Irene Poli, Pere Puigdom\`enech, Steen Rasmussen, Carlos Rodriguez-Caso, Marti Rosas-Casals, Sergio Rossi, Kashif Sadiq, Marten Scheffer, Jordi Sol\'e, Michael Tadros, Sergi Valverde, Stefanie Widder, Peter Wills and David Wolpert for useful comments, suggestions and criticisms to the ideas explored in this paper (which reflect my own views, not necessarily theirs). This  work has been supported by the Botin Foundation and by Santa Fe Institute.

 \section{references}

 \begin{enumerate}

\item
 Ackerman J (2012) The ultimate social network. Sci. Am.  June, 36-43.

\item
 Agapakis CM et al (2011) Towards a synthetic chloroplast. PLoS ONE 6: e18877.

\item
Andersson E et al (2014) Reconnecting cities to the biosphere: stewardship of green infrastructure and urban ecosystem services. AMBIO 43: 445Ð453.

\item
Armstrong R and Spiller N (2010) Living quarters. Nature 467: 916-918.

\item
Arrigo KR (2005) Marine microorganisms and global nutrient cycles. Nature 437: 349-355.

\item
Azzurro E et al (2014) External morphology explains the success of biological invasions. Ecol. Lett. doi: 10.1111/ele.12351

\item
Baker D et al (2006) Engineering Life: Building a FAB for Biology. Sci. Am. 294: 44-51.

\item
Barnes DKA (2002) Invasions by marine life on plastic debris. Nature 416: 808-809.

\item
 Barnovsky AD et al (2012) Approaching a state of shift in EarthÕs biosphere. Nature 486: 52-58.

\item
Berg P and Singer MF (1995) The recombinant DNA controversy: twenty years later. Proc Natl Acad Sci USA 92: 9011-9013.

\item
Blackburn TM et al (2011) A proposed unified framework for biological invasions. Trends Ecol. Evol. 26: 333-339.

\item
Borland AM et al (2014) Engineering crassulacean acid metabolism to improve water-use efficiency. Trends Plant Sci 19: 327-338.

\item
Brenner K et al. (2008) Engineering microbial consortia: a new frontier in synthetic biology. Trends Biotechnol 28:483Ð489

\item
Brown JH and Heske EJ (1990)  Control of a desert-grassland transition by a keystone rodent guild. Science 250: 1705-1707.

\item
Brown JH et al (2011) Energetic limits to economic growth. Bioscience 61: 19-26.

\item
Burgie ES et al (2014) Crystal structure of the photosensing module from a red/far-red light-absorbing plant phytochrome. Proc Natl Acad Sci USA 111: 10179Ð10184.

\item
Caldeira K, Bala G and Cao L (2013) The science of geoengineering. Annu. Rev. Earth Planet. Sci. 41: 231-256.

\item
Callaway S (2013) Synthetic biologists and conservationists open talks. Nature 496 : 281.

\item
Canfield DE, Glaazer Anand Falkowski PG (2010) The evolution and future of EarthÕs nitrogen cycle. Science 330: 192-196.

\item
Carnicer J et al (2011) Widespread crown condition decline, food web disruption, and amplified tree mortality with increased climate change-type drought. Proc Natl Acad Sci USA 108: 1474-1478.

\item
Cases I and de Lorenzo V (2005) Genetically modified organisms for the environment: stories of success and failure and what we have learnt from them. Intl. Microb. 8: 213-222.

\item
Cho I and Blaser MJ (2012) The human micro biome: at the interface of health and disease. Nature  Rev Genet 13: 260-270.
 
\item
Church G (2005) Let us go forth and safely multiply. Nature 438: 423.

\item
Church GM and Regis E (2012) Regenesis: How Synthetic Biology Will Reinvent Nature and Ourselves (Basic Books, New York).

\item
Costello EK (2012) The application of ecological theory toward an understanding of the human microbiome. Science 336: 1255-1262.

\item
C\'ozar A et al. (2013) Plastic debris in the open ocean. Proc Natl Acad Sci USA 111: 10239-10244.

\item
Dana GV et al (2012) Four steps to avoid a synthetic biology disaster. Nature 483: 29.

\item
Dawson TP, Jackson ST, House JI et al (2011) Beyond predictions: biodiversity conservation in a changing climate. Science 332: 53-58.

\item
DeAngelis D (1992) Dynamic of Nutrient Cycling and Food Webs. Chapman and Hall, London UK.

\item
de Lorenzo V (2008) Systems biology approaches to bioremediation. Curr Opin. Biotechnol. 19: 579-589.

\item
Dethlefsen L, McFall-Ngai M and Relman DA (2007) An ecological and evolutionary perspective on human-microbe mutualism and disease. Nature 449: 811-818.

\item
DeWall MT and Cheng DW (2011) The minimal genome: a metabolic and environmental comparison. Brief. Func. Genomics 10: 31-315.

\item
Diamond J (2002) Evolution, consequences and future of plant and animal domestication. Nature 418: 700-707.

\item
Dost‡l et al (2013) The impact of an invasive plant changes over time. Ecol Lett 16: 1277-1284.

\item
Drake JA et al Eds. (1989) Biological invasions. A global perspective (Wiley, Chichester UK).

\item
Drubin DA, Way JC and Silver PA (2007) Designing biological systems. Genes Dev. 21: 242-254.

\item
Ducat DC, Way JC and Silver PA (2011) Engineering cyanobacteria to generate high-value products. Trends Biotech. 29: 95-103.

\item
Ducat DC and Silver PA (2011) Improving carbon fixation pathways. Curr. Opin. Chem. Biol. 16: 337-344.

\item
Dufresne A et al (2003) Genome sequence of the cyanobacterium Prochlorococcus marinus SS120, a nearly minimal oxyphototrophic genome.  Proc Natl Acad Sci USA 100: 10020-10025.

\item
Dyson F (2007) Our Biotech Future. The New York Review of Books. 54(12)

\item
Elton, C.S. (1958). The Ecology of Invasions by Animals and Plants. (Methuen, London)

\item
Endy D (2005) Foundations for engineering biology. Nature 438: 449-453.

\item
Estes JA et al (2011) Trophic downgrading of planet Earth. Science 333: 301-306.

\item
Esvelt KM and Wang HH (2013) Genome-scale engineering for systems and synthetic biology. Mol Syst Biol 9:641.

\item
Evans MR (2011) Modelling ecological systems in a changing world. Phil Trans Roy Soc London B 367: 181-190.

\item
Falkowski PG, Fenchel T and Delong EF (2008) The microbial engines that drive EarthÕs biogeochemical cycles. Science 320: 1034-1039.

\item
Farre, G et al (2014) Engineering Complex Metabolic Pathways in Plants. Ann Rev Plant Biol. 65: 187-223

\item
Field CB et al (2007) Feedbacks of terrestrial ecosystems to climate change. Annu Rev Environ Resourc 32: 1-29.

\item
Foley, J. A. et al. (2003) Regime shifts in the Sahara and Sahel: interactions between ecological and climatic systems in Northern Africa. Ecosystems 6: 524Ð532.

\item
Folke C. et al. (2011) Reconnecting to the biosphere. AMBIO 40: 719-738.

\item
Fuhrman JA(1999) Marine viruses and their biogeochemical and ecological effects. Nature 399: 541Ð548.

\item
Gibson, DG. et al (2008) Complete chemical synthesis, assembly, and cloning of a Mycoplasma genitalium genome. Science 319: 1215-1220.

\item
Gregory MR (2009) Environmental implications of plastic debris in marine setting. Phil. Trans. R Soc London B 364: 2013-2025.

\item
Grimm NB (2008) Global change and the ecology of cities. Science 319: 756-760.

\item
Hobbs RJ et al (2006) Novel ecosystems: theoretical and management aspects of the new ecological world order. Global Ecol Biogeography 15: 1-7.

\item
Hobbs RJ, Higgs E and Harris JA (2009) Novel ecosystems: implications for conservation and restoration. Trends Ecol Evol 24: 599-605.

\item
Holling, C.S. (1973) Resilience and stability of ecological systems. Annu. Rev. Ecol. Syst. 4: 1Ð23.

\item
Hughes TP et al (2013) Multiscale regime shifts and planetary boundaries. Trends Ecol. Evol. 28: 389-395.

\item
Huttenhower C et al (2012) Structure, function and diversity of the healthy human microbiome. Nature 486: 207-214.

\item
Jones CG, Lawton JCG and Shachak M (1994) Organisms as ecosystem engineers. Oikos 69: 373-386.

\item
Kasting JF and Siefert JL (2002) Life and the evolution of EarthÕs atmosphere. Science 296: 1066-1068.

\item
K\'efi, S. et al. (2007) Spatial vegetation patterns and imminent desertification in Mediterranean arid ecosystems. Nature 449: 213Ð217.

\item
Keith DW (2000) Geoengineering the climate: history and prospect. Annu Rev Energy Environ. 25: 245-284.

\item
Khalil AS and Collins JJ (2010) Synthetic biology: applications come of age. Nature Rev Genetics 11: 367-379.

\item
Klausmeier CA, Lichman E, Daufresne T and Levin SA (2004) Optimal nitrogen-to-phosphorus stoichiometry of phytoplankton. Nature 429: 171-174.

\item
Ledford H (2010) Life hackers. Nature 467: 650-652.

\item
Lenton TM and van Oijen M (2002) Gaia as a complex adaptive system. Phil Trans R Soc Lond  B 357: 683-695.

\item
Lenton, T. M. et al. (2007) Effects of atmospheric dynamics and ocean resolution on bi-stability of the thermohaline circulation examined using the Grid ENabled Integrated Earth system modelling (GENIE) framework. Clim. Dyn. 29: 591Ð613.

\item
Lenton, T.M. et al. (2008) Tipping elements in the EarthÕs climate system. Proc. Natl. Acad. Sci. U.S.A. 105: 1786-1793

\item
Lenton TM and Watson A (2011) Early warning of climate tipping points. Nature Climate Change 1: 201-209.

\item
Lenton TM and Watson A (2011) Revolutions that made the Earth. Oxford U Press, Oxford UK)

\item
Levin SA et al (1997) Mathematical and Computational Challenges in Population Biology and Ecosystems Science. Science 275: 334-343.

\item
Levin SA (2000) Fragile dominion: complexity and the commons. (Basic Books, New York).

\item
Levin SA (2002) The biosphere as a complex adaptive system. Ecosystems 1: 431-436.

\item
Lin, MT et al (2014) A faster Rubisco with potential to increase photosynthesis in crops. Nature doi:10.1038/nature13776

\item
Lovelock JE (1988) The ages of Gaia. (Norton, New York)

\item
Lovelock JE and Margulis L (1974) Atmospheric homeostasis by and for the biosphere: the gaia hypothesis. Tellus 26: 2-10.

\item
Lovelock JE (1992) A numerical model for biodiversity. Phil. Trans. R. Soc. B 338: 383-391

\item
Lovelock JE and Rapley CG 2007. Ocean pipes could help the Earth to cure itself. Nature 449: 403.

\item
Lovelock, JE (2008) A geophysiologistÕs thoughts on geoengineering. Phil. Trans. R. Soc. A 36: 3883-3890.

\item
Lurgi M, L\'opez BC and Montoya JM (2012) Novel communities from climate change. Phil Trans R Soc B 367: 2913-2922.

\item
Luisi PL. (2006) The emergence of life: from chemical origins to synthetic biology. (Cambridge U. Press, Cambridge UK).

\item
Macia J, Posas F, Sol\'e RV (2012) Distributed computation: the new wave of synthetic biology devices. Trends Biotechnol 30: 342Ð349

\item
May RM (1977) Thresholds and breakpoints in ecosystems with a multiplicity of stable states. Nature 269:471-477.

\item
McCutcheon JP and Moran NA (2012)  Extreme genome reduction in symbiotic bacteria. Nature Rev Microbiol 10: 13-26.

\item
McKay CP, Toon OB and Kasting JF (1991) Making Mars habitable. Nature 352: 489-496.

\item
Michaelson J (1998) Geoengineering: A climate change Manhattan project. Stan. Env. Law J. 17: 73-140.

\item
Mittler R and Blumwald E (2010) Genetic Engineering for Modern Agriculture: Challenges and Perspectives. Annu. Rev. Plant Biol. 61: 443-462.

\item
Moe-Behrens GHG, Davis R and Haynes KA (2013) Preparing synthetic biology for the world. Frontiers Microb. 4: a5.

\item
Naeem S (1996) Species redundancy and ecosystem reliability. Conserv. Biol. 12: 39-45.

\item
New M, Liverman D, Schroder H and Anderson K (2011) Four degrees and beyond: the potential for a global temperature 
increase of four degrees and its implications. Phil rans R Soc A 369: 6-19.

\item
Nicholson WL et al (2012) Growth of Carnobacterium spp. from permafrost under low pressure, temperature, and anoxic atmosphere has implications for Earth microbes on Mars. Proc Natl Acad Sci USA 110: 666-671.

\item
Odum HT and Odum B (2003) Concepts and methods of ecological engineering. Ecol. Eng. 20: 339-361.

\item
Palumbi (2001) HumanÕs as the greatest evolutionary force. Science  293: 1786-1790.

\item
Parker IM, Simberloff D, Lonsdale WM et al (1999) Impact: toward a framework for understanding the ecological effects of invaders. Ecological Invasions 1: 3-19.

\item  
Peisajovich  SG (2012) Evolutionary synthetic biology. ACS Synth Biol 1: 199-210. 
 
\item
Pepper JW and Rosenfeld S (2012) The emerging medical ecology of the human gut micro biome. Trends Ecol Evol 27: 381-384.

\item
Peterson G, Allen CG and Holling CS (1998) Ecological Resilience, Biodiversity, and Scale. Ecosystems 1: 6Ð18.

\item
Pilson D and Prendeville HR (2004) Ecological Effects of Transgenic Crops and the Escape of Transgenes into Wild Populations. Annu. Rev. Ecol. Evol. Syst, 35: 149-174.

\item
Pimm SA (1991) The Balance of Nature. Ecological Issues in the conservation of species and communities. (Chicago U Press, Chicago).

\item
Porcar M and Peret\'o J (2014) Synthetic biology: from iGEM to the artificial cell. Springer Briefs in Biochemistry and Molecular Biology. (Springer, Dordrecht)

\item
Poulter B et al (2014) Contribution of semi-arid ecosystems to interannual variability of the global carbon cycle. Nature 509: 600-603.

\item
Purnick PEM and Weiss R (2009) The second wave of synthetic biology: from modules to systems. Nature Rev Mol Cell Bio 10: 410-422.

\item
Rasmussen S et al (2008) Protocells: bridging nonliving and living matter. MIT Press.

\item
Renda BA, Hammerling MJ and Barrick JE (2014) Engineering reduced evolutionary potential for synthetic biology. Mol. Biosys. 10: 1668-1878.

\item
Regot, S., Macia, J., Conde, N., Furukawa, K. et al. (2011) Distributed biological computation with multicellular engineered networks. Nature 469: 207-211.

\item
Rietkerk M and van de Koppel J (1997) Alternate stable states and threshold effects in semi-arid grazing systems. Oikos 79 : 69-76-

\item
Rockstrom, J. et al. (2009) A safe operating space for humanity. Nature 461: 472Ð475.

\item
Romanuk, T.M et al (2009). Predicting invasion success in complex ecological networks. Philos. Trans. R. Soc. Lond. B  364: 1743Ð1754.

\item
Roosinck MJ (2011) The good viruses: viral mutualistic symbioses. Nature Rev Microb 9: 99-108.

\item
Rothschild LJ and Mancinelli RL (2001) Life in extreme environments. Nature 409: 1092-1101.
  
\item
Ruder WC, Lu T and Collins JJ (2011) Synthetic biology moving into the clinic. Science 333: 248-252.  
 
\item
Sade N (2014) The role of tobacco aquaporin1 in improving water use efficiency, hydraulic conductivity, and yield production under salt stress. Plant Phys 152: 245-254.

\item
Sanvido O, Romeis J and Bigler F (2007) Ecological Impacts of Genetically Modified Crops: Ten Years of Field Research and Commercial Cultivation. Adv. Biochem. Eng. 107: 235-278.

\item
Sayler GS and Ripp S (2000) Field applications of genetically engineered microorganisms for bioremediation processes. Curr. Opin. Biotech. 11: 286-289.

\item
Sax DF, Stachowicz JJ, Brown JH  et al (2007) Ecological and evolutionary insights from species invasions. Trends Ecol Evol 22: 465-471.

\item
Scanlon, T. M., Caylor, K. K., Levin, S. A. and Rodriguez-Iturbe, I. (2007) Positive feedbacks promote power-law clustering of Kalahari vegetation. Nature 449: 209Ð212.

\item
Scheffer M, Carpenter S, Foley JA et al (2001)  Catastrophic shifts in ecosystems. Nature 413: 591-596.

\item
Scheffer, M. and Carpenter, S.R. (2003) Catastrophic regime shifts in ecosystems: linking theory to observation. Trends Ecol Evol 18: 648Ð656.

\item
Scheffer M, Westley F and Brock W. (2003) Slow response of societies to new problems: causes and costs. Ecosystems 6: 493-502.

\item
Scheffer M (2009) Critical transitions in nature and society (Princeton U. Press, Princeton)

\item
Schmidt M (2008) Diffusion of synthetic biology: a challenge to biosafety. Syst Synth Biol 2: 1-6.

\item
Schmidt M (2010) Xenobiology: a new form of life as the ultimate biosafety tool. Bioessays 32: 322-331.

\item
Schmidt M and de Lorenzo V (2012) Synthetic constructs in/for the environment: Managing the interplay between natural and engineered biology. FEBS Lett. 586: 2199-2206.

\item
Schneider, S. H. and Mesirow, L. E. (1976) The genesis strategy: climate and global survival 
(Plenum, New York).

\item
Schneider SH (2008) Geoengineering: could we or should we make it work? Phil. Trans. R. Soc. A 366: 3843-3862.

\item
Seastedt TR, Hobbs RJ and Suding KN (2008) Management of novel ecosystems: are novel approaches required? Front Ecol Environ 6: 547-553.

\item
Shea, K. and Chesson, P. (2002). Community ecology theory as a framework for biological invasions. Trends Ecol. Evol. 17: 170Ð176.

\item
ShouW, Ram S and Vilar JMG (2007) Synthetic cooperation in engineered yeast populations. Proc Natl Acad Sci USA 104: 1877-1882.

\item
Simberloff D and Rejm‡nek M (eds.) (2011) Encyclopedia of biological invasions. University of california Press. Berkeley CA.

\item
Simberloff D and Gibbons L (2004) Now you see them, now you donÕt! Ð population crashes of established introduced species. Biological Invasions 6: 161-172.

\item
Smillie CS et al (2011) Ecology drives a global network of gene exchange connecting the human microbiome. Nature 480: 241-244.

\item
Snow AA et al (2005) Genetically engineered organisms and the environment: current state and recommendations. Ecol. Appl. 15: 377-404.

\item
Sol\'e R and Levin SA, editors (2002) The biosphere as a complex adaptive system, theme issue. Phil. Trans. R. Soc. London B 357: 617-725.

\item
Sol\'e R and Bascompte J (2006) Self-organization in complex ecosystems. (Princeton U. Press, Princeton)

\item
Sol\'e R (2007) Scaling laws in the drier. Nature 449: 151-153.

\item
Sol\'e R, Rasmussen S and Bedau M, editors (2007a) Towards the artificial cell special issue. Phil. Trans. Royal Soc. London B 362: 1725Ð1855.

\item
Sol\'e RV, Macia J, Munteanu A and Rodriguez-Caso C (2007b) Synthetic protocell biology: from reproduction to computation. Phil. Trans. Royal Soc. London B 362: 1727Ð1739.

\item
Sol\'e R (2009) Evolution and self-assembly of protocols. Int J Biochem Cell Bio 41: 274-284.

\item
Sol\'e R (2011) Phase transitions (Princeton U. Press, Princeton)

\item
Sol\'e R and Macia J (2013) Expanding the landscape of biological computation with synthetic multicellular consortia. Nat Comput  12: 485-497.

\item
Solomon, S. et al., eds (2007) Climate Change: The Physical Science Basis. Contribution of Working Group I to the Fourth Assessment Report of the IPCC, Cambridge University Press.

\item
Steffen W. et al (2011) The Anthropocene: From Global Change to Planetary Stewardship. AMBIO 40: 739-761

\item
Stewart RIA et al (2013) Mesocosm experiments as a tool for ecological climate-change research. Adv. Ecol. Res. 48: 71-181.

\item
Strayer DL, Eviner VT, Jeschke JM and Pace ML (2006) Understanding the long-term effects of species invasions. Trends Ecol Evol 21: 645-651.

\item
Strayer DL (2012) Eight questions about invasions and ecosystem functioning. Ecol Lett 15: 1199-1210.

\item
Strayer DL, Hattala KA, Kahnle AW and Adams RD (2013) Has the Hudson River fish community recovered from the zebra mussel invasion along with its forage base? Can J Fish Aqu Sci 71: 1146-1157.

\item
Suding KN, Gross KL and Houseman GR (2004) Alternative states and positive feedbacks in restoration ecology. Trends Ecol Evol 19: 46-53.

\item
Suttle CA (2005) Viruses in the sea. Nature 437: 356-361.

\item
Szostak JW, Bartel DP and Luisi PL (2001) Synthesizing life. Nature 409: 387-390.

\item
Thornton PK, Jones PG, Ericksen PJ and Chalinor AJ (2011) Agriculture and food systems in sub-Saharan Africa in a 4¼ C+ world. 369: 117-136.

\item
Vaughan NE and Lenton TM (2011) A review of climate geoengineering proposals. Climatic change 109: 745-790.

\item
Vitousek PM et al (1997) Human domination of EarthÕs ecosystems. Science 27: 494-499.

\item
Vitousek, P.M., DÕAntonio, C.M., Loope, L.L. and Westbrooks, R. (1996). Biological invasions as global environmental change. Am. Sci. 84: 468Ð478.

\item
Wang B, Wang J, Zhang W and Meldrum DR (2012) Application of synthetic biology in cyanobacteria and algae. Frontiers Microb. 3: a344.

\item
Wang YH, Wei KY and Smolke CD (2013) Synthetic biology: advancing the design of diverse genetic systems. Annu Rev Chem Biomol Eng 4: 69-102.

\item
Ward P (2009) The Medea hypothesis. (Princeton U. Press, Princeton)

\item
Whitham TG et al (2006) A framework for community and ecosystem genetics: from genes to ecosystems. Nature Rev. Genet. 7: 511-523.

\item
Whitney SM, Houtz RL and Alonso H (2011) Advancing Our Understanding and Capacity to Engineer NatureÕs CO2-Sequestering Enzyme, Rubisco. Plant Physiol 155: 28-35.

\item
Wilhelm SW.  and Suttle CA (1999) Viruses and nutrient cycles in the sea. Bioscience 49, 781Ð788.

\item
Wilson EO (2010) The diversity of life (Harvard U Press, Harvard).

\item
Woodward G, Perkins DM and Brown LE (2010) Climate change and freshwater ecosystems: impacts across multiple levels of organization. Phil Trans Roy Soc London B 365 : 2093-2106.

\item
Wright O, Stan GB and Ellis T (2012) Building-in biosafety for synthetic biology. Microbiology 159: 1221-1225.

\item
Wright JT, Byers JE, DeVore JL and Sotka EE (2014) . Engineering or food? mechanisms of facilitation by a 
habitat-forming invasive seaweed. Ecology 159: 1221-1225.

\item
Zettler ER, Mincer TJ and Amaral-Zettler LA (2013) Life in the ÒPlastisphereÓ: Microbial communities on plastic marine debris. Env. Sci. Tech. 47: 7127-7146.

\item
Zhang Y, Ruder WC and LeDuc OR (2008) Artificial cells: building bioinspired systems using small-scale biology. Trends Biotech 26: 14-20.

\item
Zorrilla-L\'opez, U et al (2013) Engineering metabolic pathways in plants by multigene transformation. Int. J. Dev. Biol. 57: 565-576.

 \end{enumerate}

\end{document}